\documentclass[10pt,aps,prl,twocolumn,showpacs,superscriptaddress,floatfix]{revtex4-1}

\usepackage{graphicx}
\usepackage{color}\usepackage{enumerate}
\usepackage{amsmath}\usepackage{amssymb}%\usepackage{stmaryrd}\usepackage{wasysym}
\usepackage{multirow}
\usepackage{float}
\usepackage{longtable,booktabs}
\usepackage{hyperref}\usepackage{url}%Turn on when output to pdf file
\usepackage{subfigure}
\usepackage{dsfont}

\begin{document}

\title{Topologically Induced Fermion Parity Flips in Superconductor Vortices}

\author{Jeffrey C. Y. Teo}\email{jteo@virginia.edu}
\affiliation{Department of Physics, University of Virginia, Virginia 22904, USA}
\affiliation{Department of Physics and Institute for Condensed Matter Theory, University of Illinois at Urbana-Champaign, 1110 West Green Street, Urbana,  Illinois 61801-3080, USA}
\author{Mayukh Nilay Khan}
\affiliation{Department of Physics and Institute for Condensed Matter Theory, University of Illinois at Urbana-Champaign, 1110 West Green Street, Urbana,  Illinois 61801-3080, USA}
\author{Smitha Vishveshwara}
\affiliation{Department of Physics and Institute for Condensed Matter Theory, University of Illinois at Urbana-Champaign, 1110 West Green Street, Urbana,  Illinois 61801-3080, USA}

\begin{abstract}
 A highlighting feature of Majorana bound states in two-dimensional topological superconductors is that they gain a phase factor of $\pi$ upon being orbited by a vortex. This work focuses on the vortex degree of freedom itself and demonstrates that the change in the Majorana state is accompanied by a fermion parity change within the vortex. Such a parity flip is interpreted as a higher dimensional analog of the fermion parity pump mechanism in superconducting wires as well as through general topological arguments. It is demonstrated in terms of level crossings in three different situations - in i) spin-triplet paired superconductors, and in proximity-induced superconducting systems involving ii) quantum spin Hall-ferromagnet hybrids and iii) Chern insulators.
\end{abstract}

\pacs{}
\maketitle

Zero energy Majorana bound states (MBS) are exotic quasiparticles~\cite{Majorana37,Wilczek09,Beenakker11,Alicea12} that support non-local storage of quantum information and non-abelian quantum operations~\cite{Volovik99,ReadGreen,Kitaevchain,Ivanov,ChetanSimonSternFreedmanDasSarma}. They are predicted to appear as topologically protected boundary states in several condensed matter systems, such as on $p$-wave superconducting wires~\cite{Kitaevchain}, which can be realized by proximity induced superconducting strong spin-orbit coupled nanowires in magnetic fields~\cite{LutchynSauDasSarma10,Kouwenhoven12,DengYuHuangLarssonCaroffXu12,Shtrikman12,RokhinsonLiuFurdyna12,ChangManucharyanJespersenNygardMarcus12,FinckHarlingenMohseniJungLi13}, ferromagnetic atomic chains on a superconductor~\cite{NadjDrozdovLiChenJeonSeoMacDonaldBernevigYazdani14}, quantum spin Hall insulator (QSHI) - superconductor (SC) - ferrormagnet (FM) heterostructure junctions~\cite{FuKane08} and at crystalline defects in two dimensional topological superconductors~\cite{TeoHughes,BenalcazarTeoHughes}. A highlighting feature is the fermion parity switch induced in such Majorana pairs by a phase slip or the encircling of a vortex between them. In this Letter, we explore the ``back-reaction" of such a switch on the vortex and show that a unique feature emerges, namely, the internal states of the vortex itself undergo a change in parity.

More precisely, Majorana fermions are described by hermitian operators $\gamma_i$ that satisfy the Clifford relation $\{\gamma_i,\gamma_j\}=\gamma_i\gamma_j+\gamma_j\gamma_i=2\delta_{ij}$. A pair of MBS encodes a two-level system $|0\rangle$ and $|1\rangle=c^\dagger|0\rangle$, for $c=(\gamma_1+i\gamma_2)/2$ the Dirac fermion operator generated by the MBS pair. In a superconducting medium, an electronic quasiparticle acquires a $-1$ quantum phase when orbiting around a quantum vortex of magnetic flux $\phi_0/2=hc/2e$. As a  Majorana operator is a linear combination of electronic operators, a MBS also picks up a minus sign when a well separated flux vortex moves adiabatically around it. A paradox now arises from the non-local {\em fractionalization} of the electronic degree of freedom into a MBS pair. If the flux vortex only encircles one MBS, say $\gamma_2$, the Dirac fermion operator is conjugated, $c=(\gamma_1+i\gamma_2)/2\leftrightarrow c^\dagger=(\gamma_1-i\gamma_2)/2$, and the two-level system flips $|0\rangle\leftrightarrow|1\rangle$. Alternatively, the fermion parity operator, given by $(-1)^F=i\gamma_1\gamma_2$, changes sign when $\gamma_2\to-\gamma_2$. If the pair of MBS and the quantum vortex are well isolated from all other low energy modes, their total fermion parity $(-1)^{F+F_{\mbox{\tiny vortex}}}$ cannot change as tunneling of an electronic quasiparticle is thermodynamically suppressed by the excitation energy gap. The switch of fermion parity in the two-level system must therefore be compensated by a fermionic excitation at the quantum flux vortex. This manifests as a topologically protected level crossing among the Caroli-de Gennes-Matricon vortex states~\cite{CarolideGennesMatricon}. We refer to this vortex evolution as a {\em fermion parity flip}.

\begin{figure}[htbp]
\centering\includegraphics[width=0.43\textwidth]{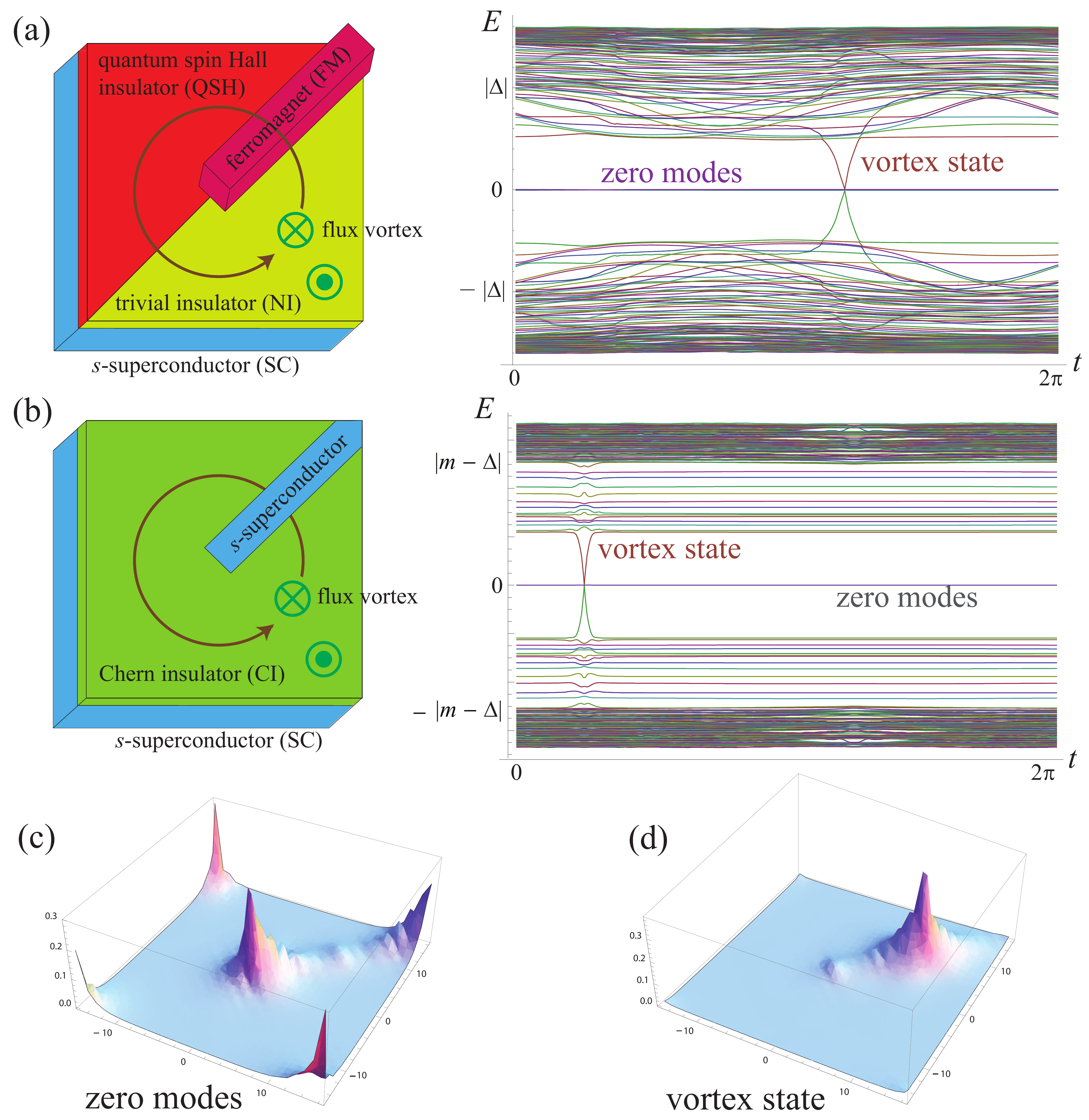}
\caption{Numerical results on a $32\times32$ periodic lattice. A single level-crossing as a vortex orbits (a) the superconducting QSH-NI-FM interface eq.\eqref{H1}, and (b) the SC trench in a Chern insulator eq.\eqref{H2}. (c) and (d) respectively show the wavefunction magnitudes of the zero energy MBS pair and the vortex state at the crossing for (b).}\label{fig1}
\end{figure}

 It is important to note that in contrast, topological phases that support Ising anyons as quantum excitations -- such as the Moore-Read fractional quantum Hall state~\cite{MooreRead} and the spinless $p+ip$ superconductor~\cite{ReadGreen,Ivanov,Kitaev06} -- cannot realize vortex parity flips. Each vortex in these systems hosts a single MBS and does not associate a local fermion parity. Or in more sophisticated language the addition of a fermion $\psi$ to an Ising anyon $\sigma$ does not change its anyon type, i.e. $\psi\times\sigma=\sigma$. Thus, there is no measurable change in fermion parity. While an even vortex with even number of Majorana's carries well-defined fermion parity $(-1)^{F_{\mbox{\tiny vortex}}}=i^n\gamma_1\ldots\gamma_{2n}$, it does not flip when the vortex orbits an Ising anyon as each vortex Majorana mode $\gamma_i$ changes sign under the cycle. Vortex fermion parity flips are therefore unique in systems where MBS bind not to vortices but to static defects.

In what follows, we first trace the conceptual origin of vortex fermion parity flip to the fermion parity pump in  one-dimensional (1D) $p$-wave superconductors~\cite{Kitaevchain,FuKaneJosephsoncurrent09,TeoKane}, which in turn is the superconducting analog of the Thouless charge pump~\cite{Thouless}. Having established the parity flip argument in 1D and the associated energy level crossing, we explore a range of instances for vortex parity flip in two dimensions (2D), each recently proposed as an exciting means of nucleating Majorana bound states.

  As an explicit 1D example, the $p$-superconducting Kitaev wire represented by the lattice Hamiltonian~\cite{Kitaevchain} \begin{align}H-\mu N=\sum_rtc^\dagger_rc_{r+1}-\mu c^\dagger_rc_r+\Delta c^\dagger_rc^\dagger_{r+1}+h.c.\label{pwire}\end{align} is topological and carries zero energy boundary MBS when the electron hopping strength $|t|$ is bigger than the chemical potential $|\mu|$. The low energy states of a superconducting ring with two weak links, one at $r=0$ and the other at $r=L/2$, are labeled by the two local fermion parities $(-1)^{F_0}$ and $(-1)^{F_{L/2}}$. When the phase of the pairing $\Delta=|\Delta|e^{i\varphi}$ winds adiabatically by $2\pi$ along a segment, say $[0,L/2]$, there is a level crossing at each of the links. This drives the vortex to an excited state after a cycle with an extra fermion, which is pumped across the bulk although there is a finite bulk pairing gap. 

\begin{figure}[htbp]
\centering\includegraphics[width=0.41\textwidth]{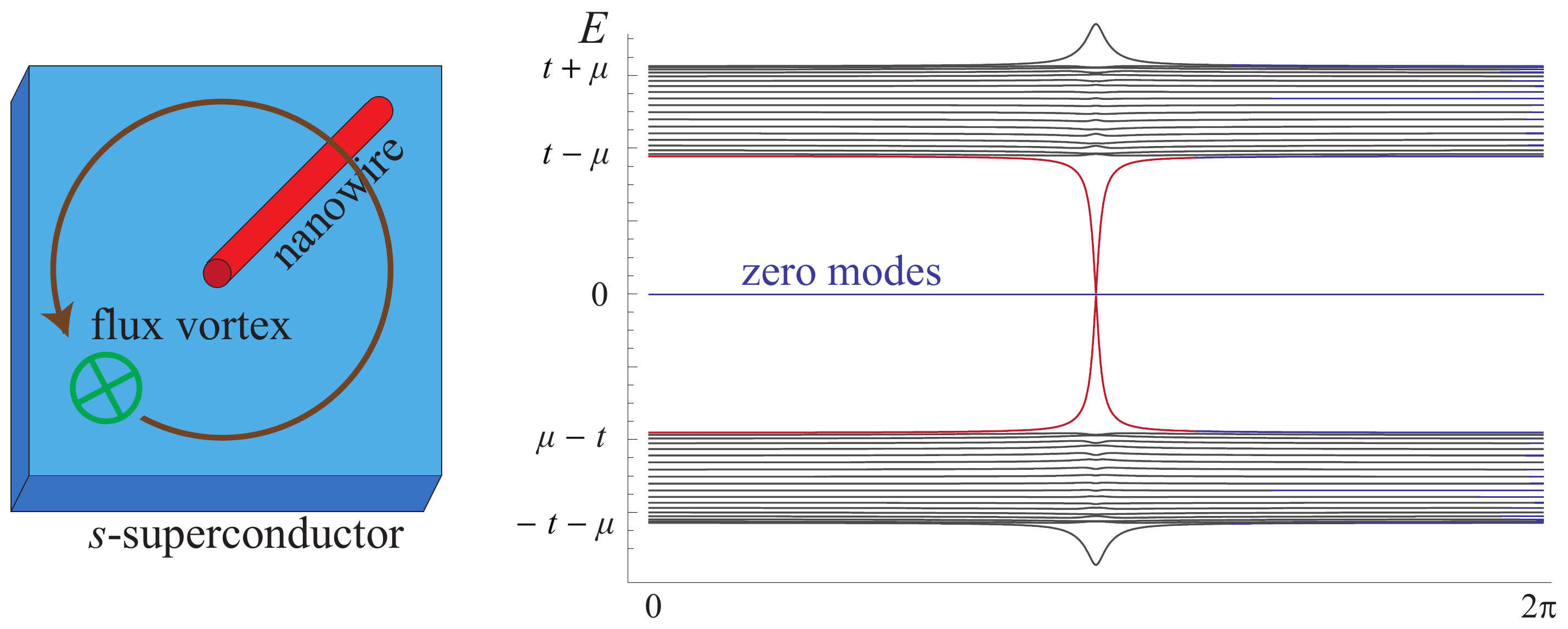}
\caption{Level crossing of a proximity induced superconducting nanowire when a $hc/2e$ flux vortex passes across.}\label{fig0}
\end{figure}
 Towards generalizing the fermion parity pump argument to higher dimension, considering passing a quantum flux vortex across a proximity induced superconducting nanowire described by Eq.\eqref{pwire}. The vortex brings spatial variation to the pairing phase $\Delta_r=|\Delta|e^{i\varphi_r}$, where $\varphi_r$ is the polar angle of site $r$ from the vortex core. Figure~\ref{fig0} shows the level crossing of a 20-site system for $t=|\Delta|=2\mu$. When the flux vortex crosses the nanowire, a Bogoliubov - de Gennes (BdG) state on the nanowire is brought down to zero energy with a wavefunction localized at the point where the vortex intersects the wire. At the same time the fermion parity of the MBS pair flips. This mimics the fermion parity pump because the pairing phase winds by $2\pi$ within the nanowire segment enclosed by the vortex trajectory. After a cycle the bulk nanowire is left with a fermionic excitation, which compensates for the parity flip of the MBS pair.

\begin{figure}[htbp]
\includegraphics[width=0.4\textwidth]{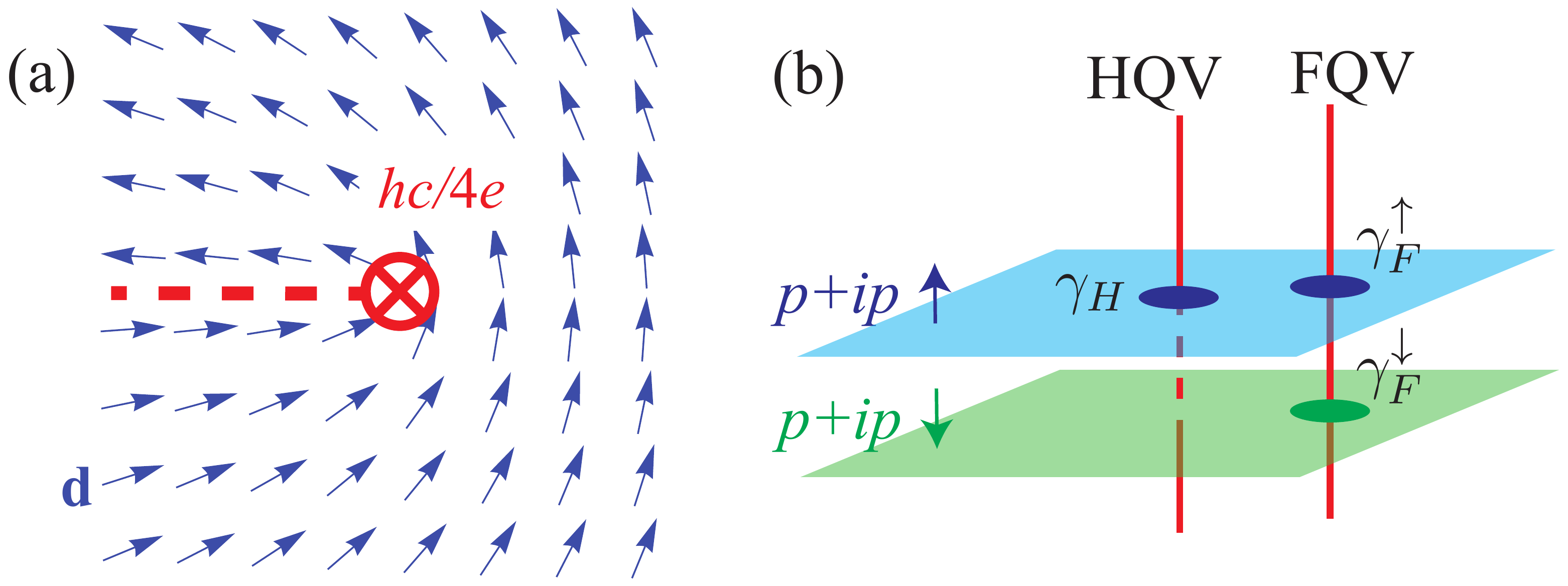}
\caption{(a) Spatial configuration of the ${\bf d}$ vector around a half-quantum vortex (HQV) of a $p+ip$ SC. (b) Zero energy Majorana modes of a half-quantum vortex (HQV) and a full quantum vortex (FQV).}\label{fig:pvortex}
\end{figure}
 Our first proposed 2D generalization of these arguments is in the context of the spin-triplet chiral superconductor having $p_x+ip_y$ pairing~\cite{Volovikbook,Leggettbook,AndersonMorel61,BalianWerthamer63,Leggett75,SigristUeda91,RiceSigrist95,LukeSigrist98,XiaKapitulnik06} celebrated for its prospects for supporting non-Abelian vortex Majorana modes~\cite{ReadGreen,Ivanov}. Assuming two spin degrees of freedom $s=\uparrow,\downarrow$, a half quantum vortex (HQV)~\cite{SalomaaVolovik85,DasSarmaNayakTewari06,ChungBluhmKim07,JangFergusonVakaryukBudakianChungGoldbartMaeno11} in such a system hosts a zero energy Majorana bound state, $\gamma_H$, at its core. The pairing $\Delta_{ss'}c_s^\dagger c_{s'}^\dagger$ associated with such a vortex can be described by $\Delta\sim|\Delta|e^{i\varphi/2}\{\partial_x-i\partial_y,({\boldsymbol\sigma}\cdot{\bf d})\sigma_y\}e^{i\varphi/2}$~\cite{BalianWerthamer63}, where the Pauli-matrices $\boldsymbol\sigma$ act on spin degree of freedom. The phase of the order parameter, $\varphi$ winds by $\pi$ around the vortex and so does  ${\bf d}$, the direction of triplet pairing (see figure~\ref{fig:pvortex}(a)). The $\pi$-winding of the ${\bf d}$ vector can be generated by the transformation $\Delta\to e^{-i\theta\sigma_z/4}\Delta e^{i\theta\sigma_z/4}$, where $\theta$ is the polar angle about the HQV. Together with the $e^{i\theta/4}$ phase from the $hc/4e$ flux, a fermion circling the vortex picks up a minus sign for $s=\uparrow$ or a trivial phase for $s=\downarrow$. Thus the zero energy Majorana mode $\gamma_H$ occupies only the spin-up sector.

Of interest here, the full quantum vortex (FQV), whose order parameter winds by $2\pi$ while the ${\mathbf d}$-vector shows no winding, contains a spectrum of bound states. Ideally, each of the spin species $s=\uparrow,\downarrow$ hosts a zero energy Majorana $\gamma_F^s$, in contrast to the HQV, where effectively the single Majorana state is attributed to one of the spin species. Unlike in the HQV, the degenerate FQV Majorana modes are fragile towards  perturbations and typically hybridize by the effective Hamiltonian $H=\lambda(-1)^F$ into a level-split $\pm\lambda$ energy pair with the two members characterized by different parities $(-1)^F=i\gamma_F^\uparrow\gamma_F^\downarrow$. Here, we specifically consider the energy splitting, $\lambda$, that arises due to a Zeeman field, ${\bf B}$. In Ref.\cite{GrosfeldSeradjehVishveshwara11}, which proposed measuring non-Abelian statistics via the Aharonov-Casher effect, it was shown that $\lambda = \mu {\bf B}\cdot {\bf d} $, where $\mu$ is the Bohr magneton. For the case where the ${\bf d}$ vector lies in-plane, in the presence of a HQV, the tunnel splitting takes the form $\mu |{\bf d}|(B_x\cos(\theta/2)+B_y\sin(\theta/2))$. As the FQV encircles the HQV as $\theta$ goes from 0 to $2\pi$, the splitting changes in magnitude and sign. There is exactly one energy crossing when $\tan(\theta/2)=-B_x/B_y$ and is protected as the FQV ground state flips fermion parity. This parity flip can be qualitatively understood by seeing both $\gamma_H$ and $\gamma_F^\uparrow$ switch signs after a cycle while $\gamma_F^\downarrow$, which lives on a separate spin sector, is unaffected by the process. As a result the Dirac operator $c=(\gamma_F^\downarrow+i\gamma_F^\uparrow)/2$ associated to the low energy FQV modes is conjugated to $c^\dagger=(\gamma_F^\downarrow-i\gamma_F^\uparrow)/2$.

For a similar reason, fermion parity flip occurs in a $(p+ip)\uparrow\times(p-ip)\downarrow$ topological superconductor (TSC). The TSC belongs to class DIII as per the Altland-Zirnbauer classification~\cite{AltlandZirnbauer97} and its topology is protected by time reversal symmetry~\cite{SchnyderRyuFurusakiLudwig08,Kitaevtable08}. Upon spin-orbit coupling, the two spin-degenerate MBS $\gamma_F^\uparrow,\gamma_F^\downarrow$ at a FQV split in energy due to the time reversal breaking magnetic flux. When the FQV circles around a HQV hosting a single MBS in one of the spin sector, there is one (or in general an odd number of) level crossing among the vortex states in the FQV.

Fermion parity flips have a general topological origin. Their presence is guaranteed by the change of sign of a single (or in general odd number of) MBS. In 2D, the BdG Hamiltonian $H({\bf k},\theta,t)$ that describes the bulk superconducting medium around the MBS varies adiabatically by the polar angle $\theta$ centered at the MBS as a function of the adiabatic/temporal parameter $t$. This class D adiabatic cycle has a non-trivial $\mathbb{Z}_2$ topological index according to the classification of topological defect~\cite{TeoKane}. The BdG Hamiltonian is topologically equivalent to a massive Dirac model \begin{align}H({\bf k},\theta,t)=k_x\Gamma_x+k_y\Gamma_y+m\Pi(\theta,t)+O({\bf k}^2)\label{Pi}\end{align} where $\Gamma_x,\Gamma_y,\Pi(\theta,t)$ are mutually anticommuting matrices and $\Gamma_x^2=\Gamma_y^2=\Pi(\theta,t)^2=1$. The mass parameter $\Pi$ lives in the classifying space $BO=\frac{O(2n)}{O(n)\times O(n)}\times\mathbb{Z}$, where $O(n)$ is the orthogonal group and $n$ is related to the number of bands in the system~\cite{FreedmanHastingsNayakQiWalkerWang}. The adiabatic evolution defines a map $(\theta,t)\to\Pi(\theta,t)$ homotopically classified by $\pi_2(BO)=\mathbb{Z}_2$, whose non-zero element characterizes a non-trivial winding and topologically protects the fermion parity flip.

The model \eqref{Pi} having a slowly varying mass term unifies 2D fermion parity flip scenarios in different systems~\cite{note1}, two more of which we now present. The first is a proximity induced superconducting (SC) interface between a quantum spin Hall insulator (QSH)~\cite{KaneMele2D1,LiuHughesQiWangZhang08,Molenkamp07,KnezRettnerYangParkinDuDuSullivan14} and a trivial normal insulator (NI). The presence of protected zero energy MBS~\cite{FuKane08} requires time reversal (TR) breaking and can be facilitated by coating an (anti)ferromagnet (FM) along the interface (see figure~\ref{fig1}(b)). We take an 8-band square lattice model \begin{align}H({\bf r},t)&=H_{\mbox{\tiny QSH-NI}}({\bf r})\otimes\tau_z\nonumber\\&\;\;\;\;+\Delta_x({\bf r},t)\tau_x+\Delta_y({\bf r},t)\tau_y+h({\bf r})\mu_y\label{H1}\\H_{\mbox{\tiny QSH-NI}}({\bf r})&=t(\sin k_x\sigma_x+\sin k_y\sigma_y)\mu_x\nonumber\\&\;\;\;\;+\left[m({\bf r})+\epsilon(2-\cos k_x-\cos k_y)\right]\mu_z\label{QSH-NI}\end{align} where $\sigma,\mu,$ and $\tau$ act on spin, orbital, and Nambu degrees of freedom, respectively. The Nambu basis is chosen to be $(c_{\uparrow,\mu},c_{\downarrow,\mu},c_{\downarrow,\mu}^\dagger,-c_{\uparrow,\mu}^\dagger)$ so that \eqref{H1} has a particle-hole symmetry $\Xi=\sigma_y\tau_yK$, for $K$ the complex conjugation operator. Eq.\eqref{QSH-NI} describes the QSH-NI interface where the mass gap $m({\bf r})$ changes sign. We assume strong SC proximity so that the induced pairing order $\Delta=\Delta_x+i\Delta_y$ is non-vanishing throughout the system. The antiferromagnet couples strongly to a strip neighborhood of the interface (see figure~\ref{fig1}) where the FM order $|h(\bf r)|$ outweights the pairing $|\Delta|$ but vanishes elsewhere.

The QSH-NI interface hosts a gapless 1D helical mode with opposite spins counter-propagating electrons. The helical mode is unstable to TR or charge conservation breaking perturbations. Its removal by magnetic field was seen in the earliest experiment of QSHI~\cite{Molenkamp07} and antiferromagnetic (FM) gapped edge was achieved in a graphene QSH state~\cite{YoungJarillo14}. Moreover, induced SC in QSH edge has been observed in HgCdTe quantum wells~\cite{MolenkampYacoby14}. 

FM and SC are competing orders along the interface and a FM-SC domain wall -- where $|h|-|\Delta|$ changes sign -- bounds a protected zero energy MBS. A pair of MBS are therefore located at the ends of the ferromagnet in figure~\ref{fig1}(a). The superconducting QSH-NI interface -- except being TR symmetric and can only be realized holographically as the edge of a 2D system -- can be treated as a Kitaev $p$-wave wire and thus carries protected boundary MBS. When a $hc/2e$ flux vortex passes across the superconducting QSH-NI interface, it is akin to traveling across a Kitaev $p$-SC where there is a single protected level-crossing among the vortex states. This signals a vortex parity flip as the vortex is excited with one extra fermion after a cycle.

We have numerically verified the vortex parity flip phenomenon via its signature level crossing by putting the model \eqref{H1} on a periodic $32\times32$ square lattice (see figure~\ref{fig1}(a)). The QSH-NI interface is located along the diagonal line and the four sides, which are sandwiched between the upper and lower triangular regions with opposite insulating mass $m$. We choose the hopping $t=m$, a uniform pairing strength $|\Delta|=0.5m$ and the antiferrormagnetic coupling $h=0.8m$ on a strip over half of the QSH-NI interface. To avoid monopole effects~\cite{monopolefootnote}, we arrange a vortex and an anti-vortex with opposite flux, depicted by ${\color{green}\boldsymbol\otimes}$ and ${\color{green}\boldsymbol\odot}$ in figure~\ref{fig1}(a). The vortices bring spatial and temporal variation to the SC pairing $\Delta({\bf r},t)=|\Delta|e^{i\varphi({\bf r},t)}$, for \begin{align}e^{i\varphi({\bf r},t)}=\frac{(z-w_1(t))\overline{(z-w_2)}}{|(z-w_1(t))(z-w_2)|}\label{phase}\end{align} where $z=x+iy$ is the complex coordinates for lattice point ${\bf r}=(x,y)$, and $w_l=x_l+iy_l$ are complex positions for the two vortex cores for $l=1,2$. The temporal dependence of \eqref{phase} comes from the circular motion of the first vortex as it orbits around a MBS when $t$ goes from 0 to $2\pi$. The second vortex is kept stationary.

Figure~\ref{fig1}(a) shows a level-crossing of vortex states and confirms the fermion parity flip. At the crossing, a unit of fermion is pumped between the vortex and the MBS pair. Unlike the $p$-wave wire case (see figure~\ref{fig0}) where the fermionic excitation is confined along the wire, here the excitation stays localized at the vortex as it moves away from the QSHI-NI interface until it is brought to the bulk bands and hybridizes with the rest of the system.

As another instance, vortex parity flip can also occur on proximity induced superconducting Chern insulators (CI)~\cite{TKNN,Haldanehoneycomb,Chang12042013} shown in figure~\ref{fig1}(b). It can be described by the 4-band BdG Hamitlonian on a square lattice \begin{align}H=&t(\sin k_x\sigma_x+\sin k_y\sigma_y)\tau_z\label{H2}\\&+[m+\epsilon(2-\cos k_x-\cos k_y)]\sigma_z+\Delta_x\tau_x+\Delta_y\tau_y\nonumber\end{align} where $\sigma,\tau$ again act on spin and Nambu degrees of freedom similar to the previous case \eqref{H1}. Without the SC pairing, \eqref{H2} describes an insulator with Chern number 1 when $-2\epsilon<m<0$. The $\sigma_z$ term is a TR breaking Zeeman coupling that competes with the induced $s$-wave pairing. We assume the pairing $|\Delta|$ is weaker than the insulating mass $|m|$ so that the 2D system is {\em not} in the chiral $p+ip$ phase~\cite{QiHughesZhang10} and a full quantum vortex does not hold a zero energy MBS.

The consequence of the bulk Chern invariant is that the CI carries a gapless chiral edge mode that propagates in a single direction~\cite{TKNN,Haldanehoneycomb}. When two uncoupled CI's with the same chirality are juxtaposed side by side, the interface bounds a pair of counter-propagating electron channels with opposite spins $\psi_{R\uparrow},\psi_{L\downarrow}$. This gapless helical interface can be gapped out by TR breaking backscattering $m\psi_{R\uparrow}^\dagger\psi_{L\downarrow}$ or U(1) breaking pairing $\Delta\psi_{R\uparrow}^\dagger\psi_{L\downarrow}^\dagger$ between the two boundaries. These orders compete and a domain wall, where $|m|-|\Delta|$ changes sign, traps a protected zero energy MBS. This can be realized by inserting a single-layer thick strongly superconducting trench in the 2D system, where $|\Delta_{\mbox{\tiny trench}}|>|m|$ (see figure~\ref{fig1}(b)).

As with the superconducting QSH-NI interface considered above, the SC trench in a CI behaves like the 1D Kitaev $p$-wire and flips the fermion parity of a passing quantum vortex. We numerically verify this by putting the SC-CI model \eqref{H2} on a $32\times32$ periodic lattice. Similar to the previous case, we arrange a vortex anti-vortex pair and consider a circular vortex trajectory around the end of the SC trench where a MBS sits. The pairing phase $\Delta({\bf r},t)=\Delta_x+i\Delta_y=|\Delta|e^{i\varphi({\bf r},t)}$ is also given by \eqref{phase}. Figure~\ref{fig1}(b) shows the adiabatic evolution of energies throughout the cycle. States between $\pm|m-\Delta|$ are vortex states and are localized at the two vortex cores. There is a single level-crossing signifying the parity flip of the vortex as it travels across the SC trench. Figure~\ref{fig1}(c) and (d) shows the localized wavefunctions of the zero energy MBS pair and the vortex state near the crossing respectively. The SC-CI setup is even more 
prefarable than the previous cases in demonstrating the vortex parity flip. Unlike the SC QSH-NI interface, the vortex excitation here stays localized at the vortex core throughout the evolution and never hybridizes with the bulk. This means that in the absence of accidental fermion poisoning, the vortex would carry a different electric charge after a complete cycle.

 Finally, we discuss how vortex parity flip has an analog in the recent context of twist defects in topological phases with anyonic symmetries~\cite{Kitaev06,Bombin,YouWen,BarkeshliJianQi,teo2013braiding,khan2014}. This new interpretation is immensely powerful and applies even to fractional MBS (or parafermions)~\cite{LindnerBergRefaelStern,ClarkeAliceaKirill,MChen,Vaezi}. Two-dimensional $s$-wave superconductors are fermion parity protected quasi-topological phases~\cite{HanssonOganesyanSondhi04,BondersonNayak13}. They have the same topological order~\cite{Wentopologicalorder90} as a $\mathbb{Z}_2$ gauge theory~\cite{Wilczekbook,BaisDrielPropitius92,Kitaev97}. A quantum vortex of $\phi=hc/2e$ takes the role of the $\mathbb{Z}_2$ flux ${\bf m}$, and an excited vortex with an addition BdG fermion $\psi$ realizes the $\mathbb{Z}_2$ charge ${\bf e}={\bf m}\times\psi$. These quasiparticles appear in the proximity induced SC QSH-NI interface, and a MBS at the SC-QSH-FM heterostructure serves as a {\em dislocation} twist defect~\cite{Kitaev06,Bombin} that switches ${\bf e}\leftrightarrow{\bf m}$ when they orbits it.

SC-CI hybrids on the other hand have a different topological order. The anyonic content is identical to a $U(1)_2$, or equivalently $SO(2)_1$, theory~\cite{khan2014}. A $hc/2e$ vortex, denoted by ${\bf m}$, supported by the SC traps a fractional charge $e^\ast=e/2$ (modulo $2e$) on the CI by the Laughlin argument~\cite{Laughlinargument} and carries semionic statistics. After a cycle around a MBS at the end of the SC trench, the vortex ${\bf m}$ is excited with an additional fermion and has different charge $-e^\ast$ (mod $2e$), and becomes the antiparticle $\overline{\bf m}={\bf m}\times\psi$. A MBS at a SC-CI heterostructure can thus be regarded as a twist defect that conjugates orbiting quasiparticles. 

In this Letter, we have shown that a highlighting feature of Majorana defect bound states in two-dimensional superconductors, namely the gaining of a phase factor of $\pi$ upon being orbited by a vortex, is necessarily accompanied by a fermion parity switch in the vortex itself. We have presented this scenario in several possible geometries relevant to recent theoretical and experimental explorations of topological systems. In principle, the parity flip would be detectable through charge sensitive measurements and would constitute not only a signature of MBS physics but also a unique parity process in and of itself.

\paragraph{Acknowledgement} We thank Taylor L. Hughes and Victor Chua for insightful discussions. This work was funded by the National Science Foundation under grant DMR 1351895-CAR (M.N.K) and DMR 0644022-CAR (S.V.). JCYT acknowledges the support from the Simons Foundation Fellowhips. We thank the support of the ICMT at the Univ. of Illinois at Urbana-Champaign.

%\bibliography{refFinal}

\begin{thebibliography}{72}
\bibitem [{not()}]{note1}%
  \BibitemOpen
  \href@noop {} {}\bibinfo {note} {The spinful $p+ip$ model can be turned into
  \eqref{Pi} by adding trivial bands and deforming the pairing variation to
  additional mass terms.}\BibitemShut {Stop}%
\bibitem [{mon()}]{monopolefootnote}%
  \BibitemOpen
  \href@noop {} {}\bibinfo {note} {A monopole would lead to delocalized zero
  modes along system boundary.}\BibitemShut {Stop}%
\end{thebibliography}
%

\end{document}